\documentclass[aps,amsfonts,pre,twocolumn,superscriptaddress,showpacs]{revtex4-1}
\usepackage{epsfig,amsmath,amssymb,bm,epsf,graphics,psfrag,verbatim,subfigure}

\def\Prob{\mathcal{P}}
\def\kNN{k}
\def\kNNN{\kappa}
\def\vR{\vec{R}}
\def\vu{\vec{u}}

\def\vq{\vec{q}}

\def\pct{\Prob_c^{\textrm{MF},\Delta}}
\def\vr{\vec{r}}
\def\dg{\Lambda}

\def\tkm{\tilde{k}_m}
\def\tkam{\tilde{\kappa}_m}

\def\dg{\Lambda}
\def\vl{\vec{l}}
\def\intnum{h}

\begin{document}
\title{Finite temperature mechanical instability in disordered lattices}
\author{Leyou Zhang}
\affiliation{Department of Physics, University of Michigan, Ann Arbor, MI 48109}
\author{Xiaoming Mao}
\affiliation{Department of Physics, University of Michigan, Ann Arbor, MI 48109}

\date{\today}

\begin{abstract}
Mechanical instability takes different forms in various ordered and disordered systems.  We study the effect of thermal fluctuations in two disordered central-force lattice models near mechanical instability: randomly diluted triangular lattice and randomly braced square lattice.  These two lattices exhibit different scalings for the emergence of rigidity at $T=0$ due to their different patterns of self stress at the transition.  Using analytic theory we show that thermal fluctuations stabilize both lattices.  In particular, the triangular lattice displays a critical regime in which the shear modulus scales as $G \sim T^{1/2}$, whereas the square lattice shows $G \sim T^{2/3}$.

\end{abstract}


\maketitle

Mechanical instability takes a rich variety of forms in systems ranging from granular matter to biological tissue and man-made structures~\cite{Maxwell1864,Calladine1978,Phillips1979,Phillips1985,Alexander1998,Thorpe2000,Wyart2005a,Broedersz2011,Lubensky2015}.  Even in lattice models where there is a remarkably simple principle -- the Maxwell's rule which compares discrete numbers of degrees of freedom and constraints~\cite{Maxwell1864} -- multiple classes of distinct behaviors arise.  Mechanical instabilities in lattice models can take the form of either mean-field~\cite{Mao2010,Ellenbroek2011,Mao2011a} or non mean-field~\cite{Jacobs1995,Jacobs1996} transitions, with continuous~\cite{Jacobs1995,Jacobs1996} or discontinuous~\cite{chubynsky,Obukhov1995,moukarzelfield,Kasiviswanathan,rivoire,Mao2013c,Barre2014,Zhang2014} changes of various elastic moduli, and in some case even with a phonon structure of nontrivial topology~\cite{Kane2014,Lubensky2015}.  One central concept in understanding these different classes of behaviors is \emph{self stress}, which describes distributions of stress on bonds in the lattice with which the net force on every site vanishes.  This is characterized by the generalized Maxwell's rule~\cite{Calladine1978}
\begin{align}
	N_0 = dN-N_b+N_s
\end{align}
where $d$ is the dimension, $N$ is the number of sites, $N_b$ is the number of central force (CF) bonds, $N_0$ is the number of floppy modes, and $N_s$ is the number of states of self stress.  According to this rule, each bond added to a lattice either removes one floppy mode or increases one state of self stress, and in the latter the bond can be considered as ``redundant.''   Moreover, states of self stress are intimately related to elastic moduli because it describes the nature of stress the lattice can sustain~\cite{Guest2003}.  
Whether a bond is redundant or not strongly depend on the architecture of the lattice~\cite{Sun2012}.  Thus a small difference in the construction of the lattice can lead to very different behaviors near mechanical instability.  Different classes of lattice models have been found to relate to different experimental systems, such as jamming of spheres~\cite{Liu2010,Mao2010}, glasses~\cite{Phillips1979,Phillips1985}, and semi-flexible polymer networks~\cite{Broedersz2011}.

Little is known about how thermal fluctuations affect mechanical instability in these different classes of models.  In general, fluctuations have been shown to change conclusions from mean-field theories about transitions dramatically, e.g., fluctuations can alter the order and scaling exponents of the transition, or even ruling out the ordered phase completely~\cite{Cardy1996}.  It has been pointed out that thermal fluctuations may stabilize under-coordinated networks in some special cases including crosslinked flexible polymer~\cite{DeGennes1979,Xing2004,Mao2009a}, tethered networks~\cite{Rubinstein1992,Barriere1995,Bowick1996,Tessier2003}, diluted lattices~\cite{Plischke1998,Dennison2013}, and jammed packings~\cite{Ikeda2012,Ikeda2013,DeGiuli2015}, 
but a generic theoretical framework for finite-temperature mechanical instability, especially in disordered systems, is still lacking.  Such generic theory will be important in understanding experimental observations of mechanical instability which are made at finite temperature.

\begin{figure}
\centerline{\includegraphics{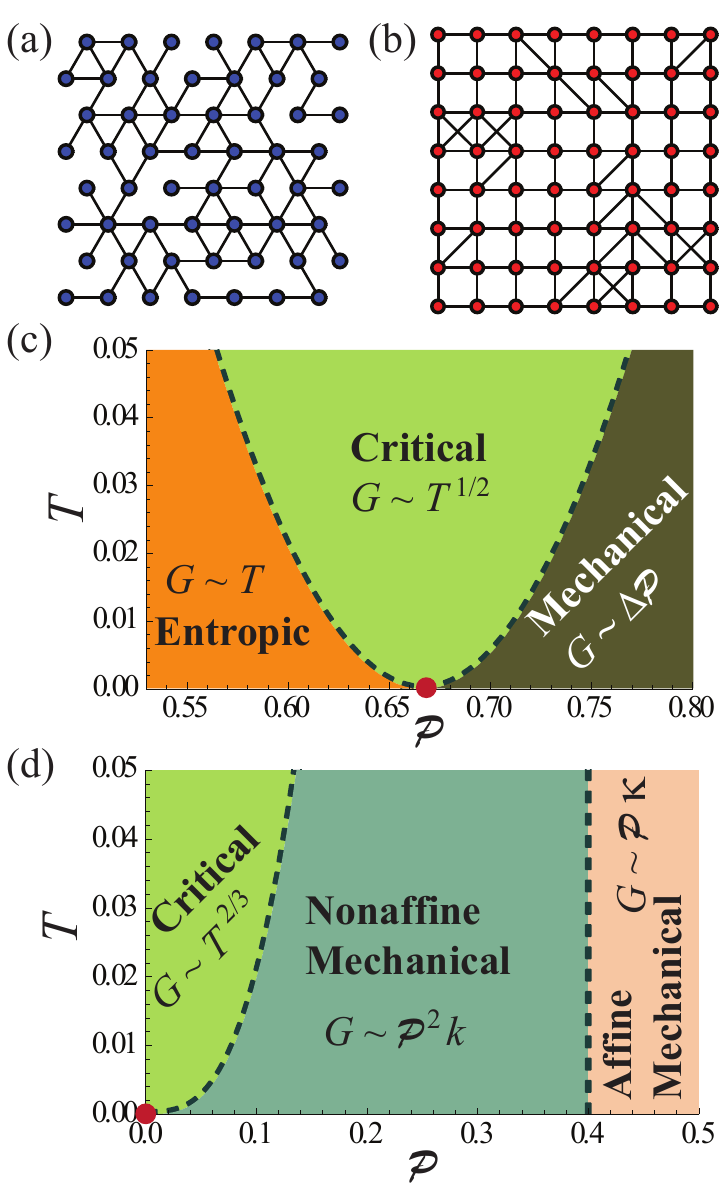}}
\caption{Randomly diluted triangular lattice (a), randomly braced square lattice (b), with their phase diagrams in the space of probability $\Prob$ for having each bond (NNN bonds in the case of square lattice) and temperature $T$ measured in units of $ka^2$ shown in (c) and (d) respectively.  In (d) NNN spring constant $\kNNN=1$.  Red dots mark the $T=0$ mechanical instability point.}
\label{FIG:pd}
\end{figure}

In this letter, we construct an analytic theory for disordered lattices near mechanical instability at finite temperature.  Our theory treats thermal fluctuations through renormalization of elastic moduli and treats lattice disorder through the coherent potential approximation (CPA)~\cite{Soven1969,Garboczi1985}.  We use this theory to discuss two CF lattice models, the randomly diluted triangular lattice~\cite{Feng1984} and the randomly braced square lattice~\cite{Mao2010}, both of which display mechanical instabilities when they pass through the CF isostatic point $\langle z \rangle =2d$ where $z$ is the coordination number.  Our main findings are summarized in the phase diagrams in Fig.~\ref{FIG:pd}.  The triangular lattice is stabilized by thermal fluctuations near the isostatic point, and it displays three regimes of elasticity, $G\sim \langle z \rangle -2d+O(T)$ (mechanical), $G\sim T^{1/2}$ (critical), and $G\sim T$ (entropic), consistent with numerical results in Ref.~\cite{Dennison2013}.  Interestingly the square lattice shows very different exponents, e.g., $G\sim T^{2/3}$ in the critical regime, as well as a nonaffine-affine crossover, originating from the sub-extensive number of states of self stress at the transition.  Our theory is readily generalizable to other models to map out finite-temperature phase diagrams for different classes of systems near mechanical instability.

\emph{Model} --- We consider two types of lattices near mechanical instability.  The first one is the randomly diluted CF triangular lattice [Fig.~\ref{FIG:pd}(a)], in which each nearest-neighbor (NN) bond is of spring constant $k$ and is present with probability $\Prob$.  Previous numerical studies~\cite{Jacobs1995} have shown that the $T=0$ rigidity transition in the triangular lattice is continuous and occur at $\Prob=0.6602\pm0.0003$ with non-mean-field exponents~\footnote{These studies were done on generic triangular lattices, in which only the topology of the connectivity is kept and the periodicity of the lattice is removed to avoid state of self stresses associated with parallel bonds.}.  The second one is the randomly braced CF square lattice [Fig.~\ref{FIG:pd}(b)], in which all NN bonds are present with spring constant $k$ and each next-nearest-neighbor (NNN) bond of spring constant $\kNNN$ is present with probability $\Prob$.  
The $T=0$ rigidity percolation of this model has been studied both numerically and analytically, revealing a transition at $\Prob=0$ with mean-field exponents and features of both first- and second-order transitions.

The Hamiltonian of both lattices can be written as
\begin{align}
	H = \sum_{\langle i,j\rangle} \eta_{ij} \frac{k_{i,j}}{2} (\vert \vR_i - \vR_j \vert -\vert \vec{l}_{ij} \vert)^2 ,
\end{align}
for a deformation that moves positions $\vr_i \to \vR_i$ for each lattice site $i$.  Here the sum is over all bonds $\langle i,j\rangle$, $\eta_{ij}=1$ if the bond is present and $0$ if not,  and $\vec{l}_{ij}=\vr_i - \vr_j$ is the vector of the bond in the reference space (assuming all bonds to be at rest length in the reference space).  In the triangular lattice $k_{ij}=\kNN$, and in the square lattice $k_{ij}=\kNN$ for the NN bonds and $k_{ij}=\kNNN$ for the NNN bonds.  

We choose to use the coherent potential approximation (CPA) for the disorder in these models.  This method, also called the effective medium theory (EMT), has been shown to work very well for mechanical properties of disordered lattices~\cite{Feng1985,He1985,TangTho1988,Mao2010,Mao2011a,Mao2013b,Mao2013c}.  The spirit of the original zero temperature CPA is that the disordered lattice can be mapped into an effective medium (EM) regular lattice in which all bonds are present with an effective medium spring constant $k_m$ as a function of dilution $\Prob$.  This mapping is determined by the self-consistency condition that if one randomly chosen bond in the EM is replaced according to the rules of the original disordered lattice (change into spring constant $k$ with probability $\Prob$ and removed with probability $1-\Prob$), the average of the perturbed phonon Green's function equals to the unperturbed EM Green's function.

In this letter we develop a version of the CPA that incorporates thermal fluctuation corrections in order to study finite-$T$ mechanical instability.  Assuming that the mapping to an EM regular lattice still holds at finite temperature, but with a temperature dependent EM spring constant $k_m(\Prob ,T)$, we characterize thermal fluctuation effects by calculating renormalized elastic moduli of the EM as follows.  Consider a deformation
\begin{align}
	\vr_i \to \vR_i = \Lambda \cdot \vr_i + \vu_i ,
\end{align}
where $i$ labels lattice sites, $\Lambda$ denote a macroscopic deformation gradient, and $\vu_i$ denote finite wavelength fluctuations around $\Lambda \cdot \vr_i$.  Free energy of the lattice at a given macroscopic deformation $\Lambda$ in finite temperature can be obtained by integrating out fluctuations $\vu_i$, which yield a vibrational entropy term.  Although the Hamiltonian contains only harmonic springs, expanding $H$ in terms of displacements generate higher order terms in displacements, and thus this entropic term depends on $\Lambda$.  
It was shown in Ref.~\cite{Mao2015} that Ward identity holds for the fluctuation corrected free energy $F(\Lambda)$.  Details of this calculation is shown in the Supplementary Information (SI).  We consider generic deformations of the form
\begin{align}\label{EQ:Lambda}
	\mathbf{\dg} = \left( \begin{array}{cc}
	1+s & t \\
	0 & 1+s
	\end{array} \right) .
\end{align}
and taking derivatives of $F$ with respect to $t$ and $s$ yields the fluctuation corrected elastic moduli.  We then obtain renormalized spring constants from these corrected elastic moduli.

Next we use the CPA procedure to calculate the perturbed Green's function as one bond is replaced
\begin{align}
	\mathbf{\mathcal{G}} = ( \tilde{D} + \mathbf{\mathcal{V}} )^{-1} = \mathbf{\mathcal{G}}^{(m)} 
	- \mathbf{\mathcal{G}}^{(m)}\cdot \mathbf{\mathcal{T}} \cdot \mathbf{\mathcal{G}}^{(m)}
\end{align}
where in the phonon Green's function $\mathbf{\mathcal{G}}$ renormalized spring constants instead of the $T=0$ ones are used.  Here $\mathbf{\mathcal{V}}$ is the perturbative potential describing changing the spring constant of a randomly chosen bond in the EM, and
% value $k_m$ to spring constants in the disordered lattice $k_s=k$ or $0$, and
\begin{align}
	\mathbf{\mathcal{T}} \equiv \mathbf{\mathcal{V}} - \mathbf{\mathcal{V}}\cdot \mathbf{\mathcal{G}}^{(m)}\cdot \mathbf{\mathcal{V}} 
	+\mathbf{\mathcal{V}}\cdot \mathbf{\mathcal{G}}^{(m)}\cdot \mathbf{\mathcal{V}}
	\cdot \mathbf{\mathcal{G}}^{(m)}\cdot \mathbf{\mathcal{V}}-\ldots
\end{align}
is the $\mathbf{\mathcal{T}}$-matrix of this perturbation.  In the case of replacing one CF bond this sum for the $\mathbf{\mathcal{T}}$-matrix can be done exactly.  
The CPA self-consistency condition 
%that the disorder average of the perturbed Green's function $\mathbf{\mathcal{G}}$ must be equal to $\mathbf{\mathcal{G}}^{(m)}$ 
can then be written in the form
\begin{align}
	\Prob \mathbf{\mathcal{T}} \vert_{k_s=k} + (1-\Prob)  \mathbf{\mathcal{T}}  \vert_{k_s=0}=0 .
\end{align}
We use this equation to determine the EM spring constants.  Details are included in the SI.

%%%%%%%%%%%%%%%%%%%%%%%%%%%%
\emph{Results}   
--- We first apply this finite temperature CPA to the diluted triangular lattices.  The renormalized elastic moduli $\tkm$ is determined by calculating shear rigidity of the EM lattice
\begin{align}\label{EQ:ST}
	G = \frac{1}{V}\frac{\partial ^2 F}{\partial t^2} \Big\vert_{t=0}
	= \frac{1}{\sqrt{3}a^2/2}\left( \frac{3}{8}k_m a^2 + \frac{c_1 T}{2} \right) ,
\end{align}
where $a$ is the lattice constant and $c_1\simeq 0.18$.  We then extract the renormalized spring constant $\tkm$ through the relation $G = (\sqrt{3}/4)\tkm$ and find
\begin{align}\label{EQ:TrigR}
	\tilde{k}_m = k_m + \frac{4 c_1 T}{3a^2} ,
\end{align}
where the last term indicates an increase of EM spring constant with $T$, characterizing fluctuation stabilization of this lattice.

\begin{figure}
\centerline{\includegraphics{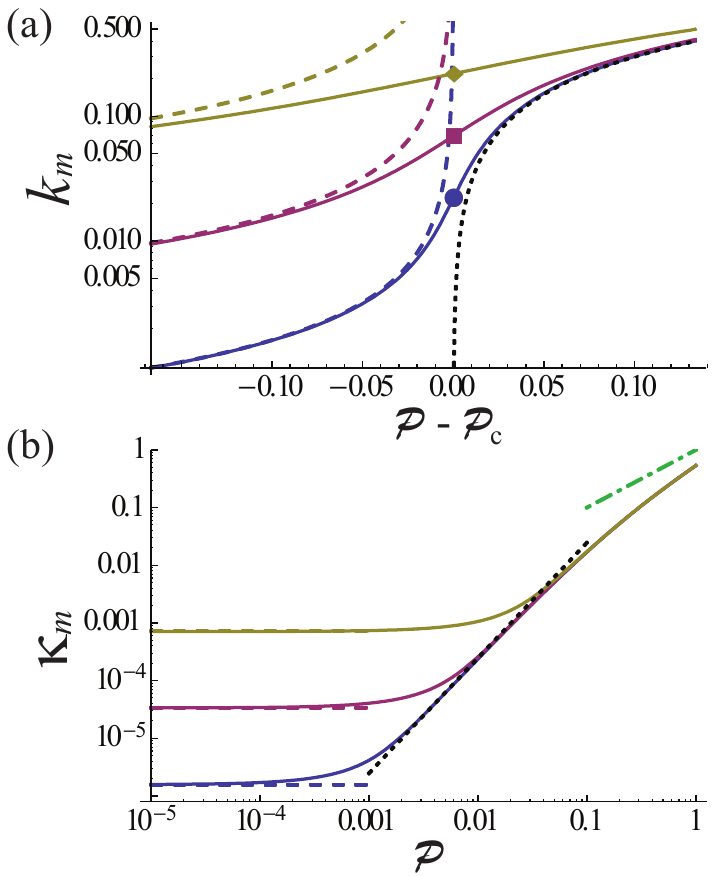}}
\caption{(a) CPA solution of $k_m$ for diluted triangular lattice as a function of $\Prob-\Prob_c$.  The solutions given by Eq.~\eqref{EQ:kmTrig} are shown in solid curves, and the asymptotic forms for the mechanical/critical/entropic regimes are shown in dotted curve/markers/dashed curves, for $T/(ka^2)=10^{-1}, 10^{-2}, 10^{-3}$ from top to bottom.  (b) CPA solution of $\kNNN_m$ for randomly braced square lattice as a function of $\Prob$.  Numerical solutions of Eq.~\eqref{EQ:SCESquare} are shown in solid curves, and the asymptotic solution for the nonaffine mechanical/affine mechanical/critical regimes are shown in dotted/dot-dashed/dashed lines, for $T/(ka^2)=10^{-4}, 10^{-6}, 10^{-8}$ from top to bottom.}
\label{FIG:G}
\end{figure}

Using this renormalized $\tkm$ we find the CPA self-consistency equation
\begin{align}
	k_m= 3\kNN\Big\lbrack \Delta \Prob+ \frac{8c_1   T}{9k_m a^2} \Big\rbrack
\end{align}
where $\Delta \Prob \equiv \Prob-\pct$ with $\pct\equiv 2/3$ being the meanfield rigidity transition point for the triangular lattice.  Taking $T\to 0$ recovers the zero temperature CPA result~\cite{Feng1985}.
%that the effective medium spring constant grows linearly with extra coordination of the network above the CF isostatic point $k_m\propto k \Delta z$, as discussed in Ref.~\cite{Feng1985}.  
In contrast, at finite $T$, with the addition of this fluctuation correction term, we find the solution
\begin{align}\label{EQ:kmTrig}
	k_m = \frac{3k \vert \Delta \Prob\vert}{2}\left\lbrack 
	1 \pm \sqrt{1 + \frac{32c_1    T}{27k a^2 (\Delta \Prob)^2}}
	  \right\rbrack ,
\end{align}
which has 3 distinct regimes
\begin{align}
	k_m = \left\{ \begin{array}{ll}
	3k\Delta \Prob & \Prob>\pct \textrm{ and }   T\ll  \frac{27ka^2 \Delta \Prob^2}{32c_1} \\
	\sqrt{8c_1 k   T/3a^2} &    T\gg \frac{27ka^2 \Delta \Prob^2}{32c_1} \\
	\frac{8c_1 k   T}{9a^2 \vert \Delta \Prob\vert} & \Prob<\pct \textrm{ and }   T\ll \frac{27ka^2 \Delta \Prob^2}{32c_1}
	\end{array} \right . ,
\end{align}
as shown in Fig.~\ref{FIG:G}(a).  
Because elastic moduli of the lattice at finite temperature are both proportional to $k_m$, these 3 regimes correspond to mechanical, critical, and entropic regimes of elasticity respectively, as shown in Fig.~\ref{FIG:pd}(c).

%%%%%%%%%%%%%%%%%%%%%%%
We apply a similar calculation to randomly braced square lattices.  In this case we determine renormalized NNN spring constant $\tkam$ by again calculating finite temperature shear modulus, which yields
\begin{align}\label{EQ:SCESquare}
	\tilde{\kNNN}_m = \kNNN_m + \frac{\pi T}{8a^2}\sqrt{\frac{k}{\kNNN_m}} .
\end{align}
Comparing to Eq.~\eqref{EQ:TrigR} we find that the renormalization of the NNN bond spring constant contains a non-analytic dependence on the ratio $k/\kNNN_m$.  This comes from the unique phonon structure of the square lattice, which exhibits lines of states of self stress and floppy phonon modes along $q_x$ and $q_y$ axis in the first Brillouin zone, leading to non-analyticity in phonon Green's functions.

The resulting finite-$T$ CPA self-consistency equation then reads
\begin{align}
	\frac{\kNNN_m}{\kNNN} = \Prob - \frac{2}{\pi} \sqrt{\frac{\kNNN_m}{k}} + \frac{T}{8a^2 \kNNN_m} .
\end{align}
Its solution can be expanded in 3 regimes too
\begin{align}
	\kNNN_m = \left\{ \begin{array}{ll}
	(\pi/2)^2 \kNN \Prob^2  & T\ll 2\pi^2 ka^2 \Prob^3 \textrm{ and }\Prob\ll \kNNN/\pi^2\\
	\Prob \kNNN &    T\ll 2\pi^2 ka^2 \Prob^3 \textrm{ and }\Prob\gg \kNNN/\pi^2  \\
	 k^{1/3}\left(\frac{\pi T}{16a^2}\right)^{2/3}\,\,   & 2\pi^2 ka^2 \Prob^3 \ll T \ll ka^2  
	\end{array} \right . ,
\end{align}
corresponding to nonaffine mechanical, affine mechanical, and critical regimes, as shown in Fig.~\ref{FIG:G}(b) and the phase diagram Fig.~\ref{FIG:pd}(d).  It is worth noting that the $T^{2/3}$ scaling agrees with the ordered lattice results in Ref.~\cite{Mao2015} with $\Prob=0$.

%%%%%%%%%%%%%%%%%%%%%%%%%%%%%%%%%%%%%%%
\emph{Discussions}   
--- In both lattice models we found that thermal fluctuations provide stabilizing effect to elasticity, even when the lattice is below the isostatic point.  Moreover, the square lattice finite-$T$ phase diagram near mechanical instability differs from that of the triangular lattice, because of the different structure of states of self stress in the square lattice.

\begin{figure}
\centerline{\includegraphics{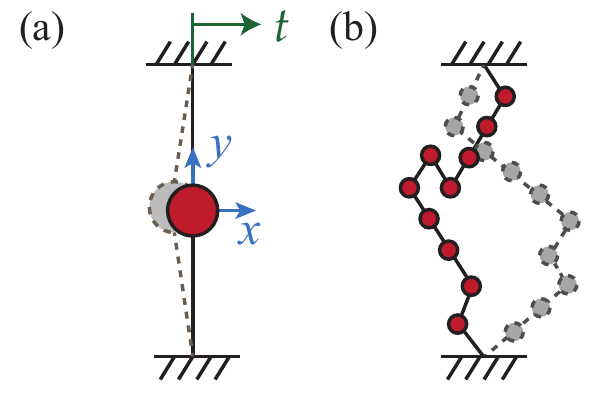}}
\caption{(a) A schematic picture explaining floppy modes near the isostatic point.  One particle (red) tethered by two springs, showing $O(T^{1/2})$ resistance to transverse deformation. One example of instantaneous fluctuation configuration is shown in gray.  (b) A schematic picture of floppy modes further below the isostatic point, in which large fluctuations (gray) cost no energy.  This leads to $O(T)$ rigidity.}
\label{FIG:oneP}
\end{figure}

There is an intuitive picture to understand the different regimes of the triangular lattice.  The mechanical regime for $\Prob>\pct$ at low $T$ is straightforward -- the lattice is stable at $T=0$ and thermal fluctuation simple provide small correction to elastic moduli.  In the critical regime, the system starts to have some modes that are of very low stiffness.  Finite temperature behavior of these modes can be understood by considering the following simple example [Fig.~\ref{FIG:oneP}(a)].  Here a particle is connected by 2 springs in the $y$ direction to 2 walls and the springs are at rest length when the particle is at $x=0, y=0$.  Assuming the 2 springs to be harmonic with force constant $k$, the Hamiltonian of the particle is to leading order
\begin{align}
	H = k\left( y^2 +\frac{x^4}{4 \ell^2} \right),
\end{align}
where $\ell$ is the rest length of the springs.  It is clear that vibrations in $x$ is a floppy mode of this simple system because its $T=0$ stiffness $\partial^2 H / \partial x^2 =0$.  Equivalently, the ``shear modulus'' of the system (resistance to transverse displacement $t$ of the top wall) is also $0$.  At finite $T$, the shear modulus can be calculated using canonical ensemble (see SI)
\begin{align}
	G = \frac{\partial^2  F}{\partial t^2} = c_2 \sqrt{k T}
\end{align}
where $c_2\simeq 0.17$.  This $T^{1/2}$ thermal stiffness for floppy modes is an intuitive way to understand the critical regime in the diluted triangular lattice, because in this regime, there is only a small number of floppy modes $U_0$ with positive coefficients for $O(U_0^4)$ terms in $H$, giving them entropic rigidity of $O(T^{1/2})$.  

When the diluted lattice is deeper below $\Prob_c$ there are more floppy modes which allow large fluctuations (flat energy landscape, no quartic term), and the corresponding physics, as shown in Fig.3b, is more like a traditional ``entropic elasticity'' picture for flexible polymers.  Thus the corresponding entropic elasticity is proportional to $T$ as in the case of polymers.

In addition, in these simple pictures and our lattice model, at finite $T$ a \emph{negative stress} is generated by thermal fluctuations, and we take fixed volume ensemble to avoid collapsing.  However, fluctuation stabilization of elasticity is not limited to the case with fixed volume.  It was shown in Ref.~\cite{Dennison2013} that with zero pressure but  self-avoidance between bonds, similar stabilization and scalings are observed.  In our theory, this correspond to allow a negative thermal expansion at finite $T$, and add additional anharmonic terms in the Hamiltonian to provide stability.  An analytic theory of this effect in ordered lattices has been discussed in Ref.~\cite{Mao2015}.

Our analytic calculations on lattice models provide a simple framework to explore fluctuation effects in more complicated systems near mechanical instability.  For example, recent studies on jamming at finite temperature revealed significant fluctuation stabilization effect below point J ($\phi<\phi_c$)~\cite{Ikeda2012,Ikeda2013,DeGiuli2015}.  It would be interesting to consider variations of our lattice models to include effects such as one-sided potential and bond-breaking to capture regimes of elasticity near point J.

\appendix

\section{Dynamical matrix and lattice free energy}
In this section we derive the free energy of an ordered lattice at given macroscopic deformations.  From the resulting free energy we get fluctuation corrected elastic moduli of the effective medium.

We write the lattice deformation as
\begin{align}
	\vr_i  \to \vR_i = \mathbf{\dg} \cdot \vr_i + \vu_i
\end{align}
where $i$ labels lattice sites, $\dg$ is the deformation gradient tensor of the (homogeneous) macroscopic deformation, and $\vu_i$ is the fluctuation deviating from the macroscopic deformation.  To capture bulk and shear moduli we take
\begin{align}\label{EQ:Lambda}
	\mathbf{\dg} = \left( \begin{array}{cc}
	1+s & t \\
	0 & 1+s
	\end{array} \right) .
\end{align}

The lattice free energy at given $\dg$ can be obtained by integrating out fluctuations $\vu$.  To do this we first expand the Hamiltonian to second order in $\vu$
\begin{align}\label{EQ:HExpa}
	H = H_0(\mathbf{\dg}) + \frac{1}{V} \sum_{q} \vu_{q} \cdot \mathbf{D}_q(\mathbf{\dg}) \cdot \vu_{-q}
	+  O((\vu)^3) ,
\end{align}
where $H_0$ is the energy of the uniformly deformed state, $\vu_q$ is the Fourier transform of $\vu_i$, and
\begin{align}\label{EQ:DML}
	\mathbf{D}_q(\mathbf{\dg}) = v_q(\mathbf{\dg}) \mathbf{I}
	+ \mathbf{\dg}\cdot \mathbf{M}_q(\mathbf{\dg}) \cdot \mathbf{\dg}^{T}
\end{align}
is the $d\times d$ dimensional ($d$ being the spatial dimension) dynamical matrix.  In this dynamical matrix the scalar $v_q$ and second rank tensor $\mathbf{M}_q$ are determined by the potentials as follows
\begin{align}
	v_q(\mathbf{\dg}) =& \sum_B 2\lbrack 1-\cos(\vq.\vl_{B})\rbrack \frac{V'_{B\dg}}{2\vert \mathbf{\dg}\cdot \vl_{B}\vert} \nonumber\\
	\mathbf{M}_q(\mathbf{\dg})  =& \sum_B 2\lbrack 1-\cos(\vq.\vl_{B})\rbrack 
	\left(\frac{V''_{B\dg}}{2} - \frac{V'_{B\dg}}{2\vert \mathbf{\dg}\cdot \vl_{B}\vert}\right) \nonumber\\
	&\quad\times\frac{\vl_{B}\vl_{B}}{\vert \mathbf{\dg}\cdot \vl_{B}\vert^2} 
\end{align}
where $\sum_B$ is a sum over bonds in one unit cell (here we only write down the simple case of one site in each unit cell), $\vl_{B}=\vr_i - \vr_j$ is the vector of this bond connecting site $i$ and $j$ in the undeformed state, and $V_{B\dg},V'_{B\dg},V''_{B\dg}$ represent the bond potential in the uniformly deformed state (of deformation gradient $\dg$) and its first and second derivatives.  There is no term linear in $\vu_{i}$ in Eq.~(\ref{EQ:HExpa}) because of the periodicity constraint.
A similar derivation has been used in Ref.~\cite{Mao2015}.

We then integrate out fluctuations $\vu$ to obtain lattice free energy,
\begin{align}
	F(\mathbf{\dg}) &= -T \ln \int \mathcal{D}\vu \, e^{-H/T} \nonumber\\
	&= H_0(\mathbf{\dg}) + \frac{T}{2} \ln \det  \mathbf{D}_q(\mathbf{\dg})    .
\end{align}
It was shown in Ref.~\cite{Mao2015} that $F(\mathbf{\dg}) $ only depend on the uniform deformation through the combination of strain tensor $\epsilon = (\mathbf{\dg}^{T}\cdot \mathbf{\dg} -\mathbf{I})$ so it keeps the same symmetry as $H_0(\mathbf{\dg})$ and thus the Ward identity is satisfied.

\section{Randomly diluted triangular lattice}
\subsection{The effective medium at finite temperature}
In this section we derive finite temperature elastic moduli of the effective medium triangular lattice, and obtain fluctuation corrections to the EM spring constant $k_m$.  

The triangular lattice unit cell contains $3$ NN bonds with
\begin{align}
	\vl_{1}=a\{1,0\}, \quad \vl_{2}=a\left\{\frac{1}{2},\frac{\sqrt{3}}{2}\right\}, \vl_{3}=a\left\{-\frac{1}{2},\frac{\sqrt{3}}{2}\right\}, 
\end{align}
where $a$ is the lattice constant.

At finite temperature the rigidity of the EM triangular lattice in response to shear deformation is given by
\begin{align}\label{EQ:ST}
	G = \frac{1}{V}\frac{\partial ^2 F}{\partial t^2} \Big\vert_{t=0}
	= \frac{1}{\sqrt{3}a^2/2}\left( \frac{3}{8}k_m a^2 + \frac{c_1 T}{2} \right)
\end{align}
where the constant
\begin{align}
	c_1 &= \frac{\partial ^2 }{\partial t^2} \Big\vert_{t=0} \textrm{Tr} \ln \mathbf{D}_q(\mathbf{\dg}) \nonumber\\
	&= \int \frac{d^2 \vq}{(2\pi)^2} \frac{\partial ^2 }{\partial t^2} \Big\vert_{t=0}\textrm{tr}\ln \mathbf{D}_q(\mathbf{\dg}) \nonumber\\
	&\simeq 0.18 ,
\end{align}
where $\textrm{Tr}$ denote a trace of the whole $Nd\times Nd$ dimensional matrix $\mathbf{D}_q$ (where $N$ is the total number of particles) and $\textrm{tr}$ denote a trace of the $d\times d$ dimensional matrix $\mathbf{D}_q$ at a given $q$.   
This finite temperature shear modulus includes contributions from both potential energy in $H_0$ and fluctuation correction term which is proportional to $T$.  

Because at $T=0$ the shear modulus is solely determined by spring constant $k_m$, at finite $T$, by identifying
\begin{align}
	G = \frac{\sqrt{3}}{4}\tilde{k}_m,
\end{align}
we can get from Eq.~\eqref{EQ:ST} the renormalized value of $k_m$,
\begin{align}\label{EQ:TrigR}
	\tilde{k}_m = k_m + \frac{4 c_1 T}{3a^2} .
\end{align}
This equation characterizes fluctuation stabilization of the lattice.

\subsection{The CPA}
In the CPA, the EM spring constant $k_m$ is determined through the following method.  We first consider picking a random bond between sites $j$ and $k$ from the lattice, and replace the spring constant of the bond from $k_m$ to $k_s$.  The scattering potential of such a replacement can be written as
\begin{align}
	\mathbf{V}_{\vq,\vq'} = (k_s-k_m)(e^{-i\vq\cdot \vr_j} -1)(e^{i\vq\cdot \vr_k} -1) \hat{\mathbf{e}}_{ij}\hat{\mathbf{e}}_{ij} ,
\end{align}
which acts as a perturbation on the dynamical matrix
\begin{align}
	\mathbf{D} \to \mathbf{D} +\mathbf{V} .
\end{align}
The resulting Green's function can be written in terms of a perturbative expansion  
\begin{align}
	\mathbf{\mathcal{G}} = ( \mathbf{D} + \mathbf{V} )^{-1} = \mathbf{\mathcal{G}}^{(m)} - \mathbf{\mathcal{G}}^{(m)}\cdot \mathbf{\mathcal{T}} \cdot \mathbf{\mathcal{G}}^{(m)}
\end{align}
where 
\begin{align}
	\mathbf{\mathcal{G}}^{(m)} = \mathbf{D}^{-1}
\end{align}
is the (unperturbed) phonon Green's function of the EM, and
\begin{align}
	\mathbf{\mathcal{T}} \equiv \mathbf{V}  - \mathbf{V} \cdot \mathbf{\mathcal{G}}^{(m)}\cdot \mathbf{V} 
	 +\mathbf{V} \cdot\mathbf{\mathcal{G}}^{(m)}\cdot \mathbf{V} \cdot \mathbf{\mathcal{G}}^{(m)}\cdot \mathbf{V} -\ldots
\end{align}
is the $\mathbf{\mathcal{T}}$-matrix of this perturbation.  The CPA self-consistency condition that the disorder average of the perturbed Green's function $\mathbf{\mathcal{G}}$ to be equal to $\mathbf{\mathcal{G}}^{(m)}$ can then be written in the form
\begin{align}\label{EQ:SCE}
	\Prob \mathbf{\mathcal{T}}\vert_{k_s=k} + (1-\Prob) \mathbf{\mathcal{T}} \vert_{k_s=0}=0 .
\end{align}
This equation determines the EM spring constant $k_m(\Prob,T)$.  

In the case of replacing one CF bond this sum for the $\mathbf{\mathcal{T}}$-matrix can be done exactly, because
\begin{align}
	&\mathbf{V}_{\vq,\vec{p}} \cdot \mathbf{\mathcal{G}}^{(m)}_{\vec{p}}\cdot \mathbf{V}_{\vec{p},\vq'} \nonumber\\
	=&  (k_s - k_m) \mathbf{V}_{\vq,\vq'} \int \frac{d^2 \vq}{(2\pi)^2} 2\lbrack 1-\cos(\vec{p}.\vl_{B})\rbrack 
	\hat{\mathbf{e}}_{B} \cdot \mathbf{\mathcal{G}}^{(m)}_{\vec{p}} \cdot \hat{\mathbf{e}}_{B} \nonumber\\
	=& (k_s - k_m) \mathbf{V}_{\vq,\vq'} \frac{\intnum}{k_m}
\end{align}
where $\hat{\mathbf{e}}_{B}\equiv \vl_B/\vert \vl_B\vert$ is the unit vector along bond $B$, and 
in the last line we write the integral as $\intnum/k_m$ where $\intnum$ is a dimensionless number.  Using this the $\mathbf{\mathcal{T}}$-matrix can be written as
\begin{align}
	\mathbf{\mathcal{T}} = \left(1+\frac{k_s - k_m}{k_m} \intnum  \right)^{-1}\mathbf{V} .
\end{align}
Plugging this into the self-consistency equation~\eqref{EQ:SCE} we have
\begin{align}\label{EQ:kmTrig}
	\frac{k_m}{k} = \frac{\Prob -\intnum}{1-\intnum}.
\end{align}

At $T=0$ we use the bare value $k_m$ in the EM Green's function $\mathbf{\mathcal{G}}^{(m)}$, and the resulting integral is
\begin{align}
	\intnum(0) = 2/3,
\end{align}
which give the meanfield $T=0$ rigidity percolation point for the triangular lattice $\pct = 2/3$.  At $T>0$ the renormalized value as in Eq.~\eqref{EQ:TrigR} should be used in the Green's function and we have
\begin{align}
	\intnum(T) = \frac{2}{3} \frac{k_m}{\tilde{k}_m} 
\end{align}
and the self-consistency equation becomes
\begin{align}
	\frac{k_m}{k} = 3\left( \Prob - \frac{2}{3} +\frac{8c_1 T}{9k_m a^2} \right) ,
\end{align}
which leads to different regimes of entropic elasticity as discussed in the main text.

\section{Randomly braced square lattice}
In this model the NN bonds are all present so the NN spring constant keeps the same value $\kNN$, and the CPA refers to mapping randomly placed NNN bonds to uniform NNN bonds on every site with an EM spring constant $\kNNN_m$.

Each unit cell in the EM has $2$ NN bonds and $2$ NNN bonds, with vectors
\begin{align}
	\vl_1 &= a\{1,0\}, \quad \vl_2 = a\{0,1\} \nonumber\\
	\vl_3 &= a\{1,1\}, \quad \vl_4 = a\{-1,1\} .
\end{align}

In order to extract the renormalized $\kNNN_m$ we need to consider a simple shear of the square lattice, which to leading order only deforms the NNN bonds.  Similar to the calculation for the triangular lattice we have
\begin{align}\label{EQ:STSq}
	G = \frac{1}{V}\frac{\partial ^2 F}{\partial t^2} \Big\vert_{t=0}
	= \frac{1}{a^2}\left( \kNNN_m a^2 + \frac{\pi T}{8}\sqrt{\frac{k}{\kNNN_m}} \right) .
\end{align}
In contrast to the fluctuation correction in the triangular lattice case as in Eq.~\eqref{EQ:ST}, here the non-analytic dependence on $k$ and $\kNNN_m$ comes from the sub-extensive number of floppy modes along the $q_x$ and $q_y$ directions in momentum space.  These floppy modes provide a divergent fluctuation correction to elastic moduli as the lattice approaches instability $\kNNN_m \to 0$.  The integral involving these floppy modes has been discussed in detail in Ref.~\cite{Mao2010}.  
As a result the renormalized NNN spring constant is
\begin{align}
	\tilde{\kNNN}_m = \kNNN_m + \frac{\pi T}{8a^2}\sqrt{\frac{k}{\kNNN_m}} .
\end{align}
Rigorously speaking, the NN spring constant $k$ also get renormalized by fluctuation, but because in this problem $k$ is always large its renormalization does not have any significant effect on stability.

In the CPA of the randomly braced square lattice we replace one arbitrary NNN bond.  The calculation for the perturbed Green's function follows a similar procedure as we did for the triangular lattice.  The self-consistency equation takes a similar form
\begin{align}
	\frac{\kNNN_m}{\kNNN} = \frac{\Prob - h}{1-h}
\end{align}
with 
\begin{align}
	h \equiv \kNNN_m \int \frac{d^2 \vq}{(2\pi)^2} 2\lbrack 1-\cos(\vec{p}.\vl_{B})\rbrack 
	\hat{\mathbf{e}}_{B} \cdot \mathbf{\mathcal{G}}^{(m)}_{\vec{p}} \cdot \hat{\mathbf{e}}_{B} 
\end{align}
where $B$ represent an arbitrary NNN bond.

The function $h$ for NNN bonds in the square lattice behaves rather differently from $h$ in the triangular lattice as we discussed above.  At $T=0$, instead of approaching a constant we have
\begin{align}
	h = \frac{2}{\pi} \sqrt{\frac{\kNNN_m}{k}} ,
\end{align}
with the non-analytic dependence coming from the sub-extensive number of floppy modes.  This $T=0$ solution has been discussed in Ref.~\cite{Mao2010}.  At $T>0$, we replace $\kNNN_m \to \tilde{\kNNN}_m$ in the phonon Green's function $\mathbf{\mathcal{G}}^{(m)}$ (not in $h$ which has an extra prefactor of $\kNNN_m$) and expand at low temperature.  This lead to the finite temperature CPA self-consistency equation
\begin{align}
	\frac{\kNNN_m}{\kNNN} = \Prob - \frac{2}{\pi} \sqrt{\frac{\kNNN_m}{k}} + \frac{T}{8a^2 \kNNN_m} .
\end{align}

\section{One particle and two springs}
In this section we discuss the simple picture of one particle tethered between two walls by two collinear springs.  To calculate shear modulus, we introduce a transverse displacement of the two walls, $t/2$ on the top and $-t/2$ on the bottom.  (This is equivalent to fixing one wall and only displace the other one by $t$ but that way one need to expand around new equilibrium position of the particle.)  The Hamiltonian is then
\begin{align}
	H = \frac{k}{2} \left\lbrack \left(y+\frac{(x+t/2)^2}{2\ell}\right)^2
	+ \left(y-\frac{(x-t/2)^2}{2\ell}\right)^2
	  \right\rbrack .
\end{align}
One can then calculate the canonical partition function at fixed $T$ by integrating over particle position $(x,y)$,
\begin{align}
	Z &= \int dxdy \, e^{-H/T} \nonumber\\
	&= \int dx \, e^{-\frac{k}{T}(\frac{x^4}{4\ell^2} + \frac{t^2 x^2}{8\ell^2}
	+\frac{t^4}{64 \ell^2}} )
	\int dy\, e^{-\frac{k}{T} (y+\frac{tx}{2\ell})^2} \nonumber\\
	&= \frac{\vert t\vert}{2}\sqrt{\frac{\pi T}{k}} K_{1/4} \left( \frac{k t^4}{128 \ell^2 T}\right) ,
\end{align}
%The integral over $y$ is simply Gaussian, whereas the integral over $x$ yields modified Bessel function of the second kind.  
where $K$ represent modified Bessel function of the second kind.  
Taking derivatives of the resulting free energy $F=-T \ln Z$ with respect to $t$ gives
\begin{align}
	G = \frac{\partial^2  F}{\partial t^2} \Big\vert_{t=0}= 
	\frac{\Gamma(3/4)}{2\Gamma(1/4)} \sqrt{k T}
	\simeq 0.17 \sqrt{k T} .
\end{align}
So the system exhibit a shear modulus that is proportional to $\sqrt{T}$.

%\bibliographystyle{aipauth4-1}
%\bibliography{isostaticity3}

%

\end{document}